\documentclass{article}

\usepackage[]{graphicx}
\usepackage[]{color}
\usepackage{alltt}
\usepackage{multirow}
\usepackage{booktabs}

\usepackage[numbers]{natbib} 
\usepackage{url}

\title{Data Mining Cultural Aspects of Social Media Marketing}
\author{Ronald Hochreiter \and Christoph Waldhauser}
\date{April 2014}

\begin{document}

\maketitle

\begin{abstract}
For marketing to function in a globalized world it must respect a diverse set of local cultures. With marketing efforts extending to social media platforms, the crossing of cultural boundaries can happen in an instant. In this paper we examine how culture influences the popularity of marketing messages in social media platforms. Text mining, automated translation and sentiment analysis contribute largely to our research. From our analysis of 400 posts on the localized Google+ pages of German car brands in Germany and the US, we conclude that posting time and emotions are important predictors for reshare counts.
\end{abstract}

\section{Introduction}

To a large part, marketing can be summarized as giving consumers what they want \cite{vargo2004evolving,merz2009evolving}. In the right contexts, a proven method is to do this informally \cite{chevalier2006effect}. While informal, heard it through the grapevine communication channels have always been important to marketing niche products and in conquering new markets, social media enables marketers to use word of mouth propagation for more established mainstream products as well \cite{rajagopal2013grapevine,kitchen2013dominant}. Since word of mouth is a very powerful vehicle to transport marketing messages \cite{kozinets2008wisdom}, marketers seek to harness the power of social media to enlist users not only as consumers but as propagators and endorsers of products \cite{sanchez2013social}, to e.g. make them partners in the co-creation process of a brand. As Kitchen \cite{kitchen2013dominant} demonstrates, social media can be used very cost effectively, enabling marketers to reach millions of users with only 
a negligible amount of resources.

It is therefore tempting to use social media marketing efforts to spread across traditional borders and reach for new markets. However, there is a risk associated with this: communication needs to be careful when crossing borders venturing into the realms of other cultures. 

A lot has been written about the need for culturally-aware communication and the management of global brands in a globalized economy \cite{alden1999brand,kotler2006b2b,matthiesen2005hugo,roth1995effects}. It is generally recommended \cite{torelli2013globalization}, to develop culturally similar markets when moving one's brand abroad. This is, because culturally accurate (i.e. functioning) translations of marketing and branding messages are very complex to produce. In this paper we seek to examine how the access to word of mouth propagation changes across cultures. To answer this question we employed data and sentiment mining techniques to the social media posts of two brands (BMW and Audi) in two countries (Germany and the US) and compared the factors that contributed to these posts being endorsed by users.

This paper is organized as follows. We will first review the literature on social media, our target platform Google+ and how marketing is done there. We then turn our attention to cultural aspects of communication and marketing. We complete this paper's theoretical part with a concise review of sentiment mining methodology. We then present the method behind our data harvest/generation and introduce the statistical models we optimized. Finally, we discuss our findings and close with some concluding remarks that point to further research.

\section{Social Media Marketing} 

In this section we focus on a rather novel arena of marketing: social media marketing. To this end we will give a short run-down on social media and then introduce a recently becoming increasingly popular social media platform: Google+ \cite{raab2011analytics,ganahl2013social}. Because of its comparably young age, Google+ has not yet received widespread attention in academia, apart from its technical aspects \cite{viegas2013google+}. This is perhaps due to its somewhat differing implementation of classical social media.

Social media and online social networks refer to a rather new phenomenon in human, internet-based communication. Diverging from a dogma that had been valid for about two decades, users themselves started out using blogs to regularly provide content themselves. While blogs were and still are appealing to users interested in writing longer texts, social media as a mass phenomenon took off only after the introduction of communities centered around profiles, frequent status updates and shared content creation. Today, leading examples of social media platforms are Facebook, Twitter, vKontakte and Google+, to name but a few. While the implementation details differ from a technical point of view, there is a common theoretical framework.

In the center of social media are user following relationships. They can be thought of as (directed) graphs linking up users. From a network theoretical point of view, this graph has small world properties and has a node distribution that follows a power law \cite{kwak2010what,java2007we}, thus being very similar to actual, offline human behavior. This relationship is called following and implies that messages sent by a user will be pushed into the stream of news all of her followers receive. While some social media platforms require reciprocal fellowship relations (e.g.\ Facebook, LinkedIn), others don't (e.g.\ Google+).

Once a message gets pushed to a user, that user can decide on how to further treat the message. Besides the obvious ignoring, a user has two levels of endorsement to choose from: liking and resharing. Endorsements are then pushed further downstream the user's network. The lesser form of endorsement, liking, does usually not contain the endorsed message but only the fact that a message from the original author was endorsed and a link to that message. User interfaces will also show likes not as prominently as reshares. With reshares the entire original message gets pushed into the streams of the user's followers, just as if the user had posted the message herself.

Another corner stone of social media are profile pages. There individuals and companies can maintain a presence with information related to them \cite{sago2013factors}. Most platforms distinguish between profiles for humans and pages for businesses and brands. However, conceptually, they are the same.

Google's implementation of a social media platform, termed Google+ and albeit being quite young, is 
increasing rapidly in popularity \cite{raab2011analytics,ganahl2013social,sago2013factors}. Here, following relationships are organized in circles, that act like address books or friend lists and allow for a more targeted resharing. While implementing a lesser form of endorsement (+1ing), it is not quite clear how that endorsement functions. For one, it is used in Google's main business of information retrieval by allowing users to discover search results that have been endorsed by the people they follow. +1s also contribute to Google internal popularity metrics, i.e. recommending the circling of possible contacts based on similar +1ing behavior. Finally, +1ed posts \emph{might} show up in a user's stream, if Google deems them algorithmically interesting. However, the exact mechanics have not yet been published by Google.

Using social media for marketing purposes is an obvious choice: being able to interact with consumers in channels usually associated with friends and interesting people we follow, makes for a very attractive marketing context. This turns marketing presence into consumer activation \cite{kitchen2013dominant}. A priceless achievement. Added to this is the possibility of marketing messages being reshared, thus profiting from traditional offline word-of-mouth benefits \cite{goyal2012facebook,mcnair2011introduction,dennhardt2012user}.

Under these terms, marketers must aim to maximize the resharing of their messages. It is useful to think of the individual user as a filter through which a message must pass in order to reach further into the network of users \cite{rogers2010diffusion,shannon2002mathematical,vos2013social}. A lot of research effort has been put into finding predictors for the expected reshare count of a message \cite{kempe2003maximizing,hochreiter2013stochastic}. Message sentiment is identified by \cite{cha2010measuring,stieglitz2012political} as crucial predictors for resharing counts as is message length \cite{stieglitz2012political}. Temporal proximity and time of day are factors named in \cite{macskassy2011people}'s contribution.   

\section{Cultural Aspects of Marketing}

Starting with Geert Hofstede's massive and groundbreaking survey of intercultural communication \cite{hofstede1984hofstede}, cultural aspects quickly became important when adapting marketing messages to local audiences of consumers \cite{taras2010examining,nakata2009beyond}. 

\begin{table}
\caption{\label{tab:hof}Scores in Hofstede's cultural dimensions for Germany and the US \cite{hofstede2010cultures}.}
\centering
\begin{tabular}{lrr}
\toprule
Dimension & Germany & US \\
\midrule
Power distance & 35 & 40\\
Individuality & 67 & 91\\
Masculinity & 66 & 62\\
Uncertainty avoidance & 65 & 46\\
\bottomrule
\end{tabular}
\end{table}

Applying Hofstede's insights to marketing, \cite{rajagopal2013grapevine,kitchen2013dominant,torelli2013globalization} come to the conclusion that marketing messages must be adapted to the cultural expectations of consumers for them to function. When looking at marketing in social media, this should also be true. Consumers organized into localized brand pages should differ in their preferences on marketing messages, just as they would differ in offline communication. From Hofstede's difference matrix, it is to be expected to find different marketing message preferences in the dimensions of \emph{Individuality} and \emph{Uncertainty avoidance} for our cases of Germany and the US. These differences should express themselves in terms of the importance of the filtering criteria per country, as introduced above.

In the next section we will review how sentiment mining can help in providing message characteristics for discriminating along the lines of cultural preferences. 

\subsection{Sentiment Mining}

Sentiment analysis and opinion mining are sub-fields of the area of text mining, see e.g. \cite{liu2012sentiment}. It is a classification task and represents the computational study of sentiments, subjectivity, appraisal, and emotions  expressed in text. Companies usually spend huge amounts of money to find consumer opinions using consultants, surveys, and focus groups. A cleverly implemented sentiment mining tool supports such companies to save money. 

An opinion is simply a positive or negative sentiment, view, attitude, emotion, or appraisal about an entity or an aspect of the entity \cite{huliu2004opinion} from an opinion holder \cite{bethard2004opinion}. The sentiment orientation (sentiment polarity) of an opinion can be positive, negative, or neutral (no opinion). 

Besides in the (sentiment) polarity of a respective Google+ post, we are interested in the emotionality of a posting. However there is no agreed set of basic emotions of people among researchers. Parrott \cite{parrott2001emotions} identifies six main emotions, i.e. love, joy, surprise, anger, sadness, and fear, whereby the strengths of opinions and sentiments are sometimes related to certain emotions, e.g., joy, anger. We apply the R package \texttt{sentiment} \cite{rsentiment} to compute the following emotions: anger, disgust, fear, joy, sadness, and surprise.

\subsection{Data \& Methods}

The generation of the data set used in this analysis started out with the identification and designation of Google+ company pages. We settled for two German companies in the automotive industry, BMW and Audi, as German car brands in different countries make a good middle ground between niche and mainstream markets \cite{guerzoni2014automotive}.

Measuring culture of a company or any individual can be elusive. It is possible to use proxies like country of residence or nationality to locate an individual culturally \cite{schuman2009impact}. We are following a similar path by looking at the companies' respective local Google+ pages. We therefore chose the localized variants of the Google+ pages of Audi and BMW for Germany and the US.\footnote{Even though Audi and BMW are both German brands, their main Google+ pages are international fronts. So both localized version are comparable in catering to local audiences.}

The following Table \ref{tab:followers} shows the popularities of the respective companies' local Google+ pages. Google offers two metrics to measure the popularity of a page. One is the circle count, i.e. the number of people that have subscribed to push updates of that page. The other one is the +1 count that aggregates the circle count with people +1ing the page or interacting with it in other ways.\footnote{C.f.\ \url{http://googleblog.blogspot.co.at/2011/12/google-few-big-improvements-before-new.html}.}

\begin{table}
\caption{\label{tab:followers}Popularity of local G+ pages of German car brands. All data was retrieved on January 8\textsuperscript{th}, 2014.}
\centering
\begin{tabular}{lllrr}
\toprule
\textbf{Brand} & \textbf{Country} & \textbf{Page ID} & \textbf{Circle count} & \textbf{+1 count} \\
\midrule
\multirow{2}{*}{Audi} & Germany & audide & 67486 & 94337 \\
 & US & AudiUSA & 1623374 & 2116081\\
 \multirow{2}{*}{BMW} & Germany & BMWDeutschland & 84429 & 113853 \\
 & US & BMWUSA & 39946 & 84042 \\
 \bottomrule
\end{tabular}
\end{table}

In order to build up a data base of marketing messages, the last 100 posts of each of these pages were retrieved. Due to different posting frequencies, the extent of the data's retrospection varies. Table \ref{tab:ages} summarizes these differences. From these figures it is obvious, that English language content for the US versions of the G+ presences is provided far more frequently than for their German counterparts.

\begin{table}
\caption{\label{tab:ages}Average posting frequencies and extents of the retrospection for harvested data.}
\centering
\begin{tabular}{llrr}
\toprule
\textbf{Brand} & \textbf{Country} & \textbf{Posts/day} & \textbf{Start date} \\
\midrule
\multirow{2}{*}{Audi} & Germany & 0.3112 & 2013-2-20\\
 & US & 0.7133 & 2013-8-20 \\
 \multirow{2}{*}{BMW} & Germany & 0.5124 & 2013-6-26 \\
 & US & 0.763 & 2013-8-29 \\
\bottomrule
\end{tabular}
\end{table}

In the following we will describe the data set in greater detail. Table \ref{tab:sumTab} gives an overview of directly measured message properties. The variable \emph{Age} describes how many days in the past the message had been posted. This obviously influences the number of reshares a message received or could receive. Message length is a property that in the past has often been used successfully to predict message reshare counts \cite{hochreiter2013stochastic}. 

\begin{table}[ht]
\centering
\caption{Ranges, means and standard deviations of recorded message properties.} 
\label{tab:sumTab}
\begin{tabular}{lrrrr}
  \toprule
Variable & Min & Mean & Max & SD \\ 
  \midrule
Age & 3.45 & 98.70 & 325.33 & 70.59 \\ 
  Number of reshares & 0.00 & 11.21 & 101.00 & 13.38 \\ 
  Number of comments & 0.00 & 12.23 & 173.00 & 18.86 \\ 
  Number of +1s & 12.00 & 170.75 & 928.00 & 150.77 \\ 
  Message length & 0.00 & 262.88 & 1748.00 & 388.23 \\ 
   \bottomrule
\end{tabular}
\end{table}

As mentioned above, another aspect that -- potentially -- influences the frequency with which a message is being reshared is the time of day of the original post and the day in the week. For this analysis we recorded the date and time of the original post (and converted the UTC timezone reported by Google to EST for the US and CET for Germany) and distilled the day of week from it as well as the discrete factorization of the time of day part that is given in Table \ref{tab:todf}.

\begin{table}[ht]
\centering
\caption{Distribution of time of day periods.} 
\label{tab:todf}
\begin{tabular}{lr}
  \toprule
Period & Frequency \\ 
  \midrule
6 a.m. -- 9 a.m. &   3 \\ 
  9 a.m. -- 12 p.m. &  53 \\ 
  12 p.m. -- 5 p.m. & 240 \\ 
  5 p.m. -- 8 p.m. &  63 \\ 
  8 p.m. -- 1 a.m. &  40 \\ 
  1 a.m. -- 6 a.m. &   1 \\ 
   \bottomrule
\end{tabular}
\end{table}

While we did record the number of comments and +1s every message received, these forms of interaction and propagation are not considered any further in this analysis for their ambiguous meaning in marketing contexts as detailed above.

Aside from these directly measured attributes of a message, we also computed every message's sentiment. German language messages (from the local German G+ pages) were automatically translated using the Google Translate API\footnote{C.f.\ \url{https://developers.google.com/translate/}.}. While automatic translation might not always work perfectly, wrong translations would only increase statistical variation and thus lead to conservative hypothesis test results. Therefore, effects that can be found using automatic translation are very likely to indeed occur in the sample population \cite{brown1993mathematics}.

All computations (and indeed the harvesting for that matter) were done using R version 3.0.2 \cite{rcore} and the \texttt{plusser} extension package \cite{plusser}.  Computation and visualization was aided by the R packages \texttt{ggplot2} \cite{rggplot} and \texttt{MASS} \cite{mass}. As detailed above, sentiment analysis extends into two distinct areas: polarity and emotions. The used software package offers to compute the positive---negative ratio, the ratio of the absolute log-likelihoods of the message expressing a positive or negative sentiment, to measure polarity of a message. A value of 1 indicates a neutral statement, while values smaller than 1 point towards a negative statement.

Measuring the involved emotions is a little bit more complicated. Here, six dimensions of frequently occurring emotions have been identified and for every message the log-likelihood of it reflecting these emotions is computed. Table \ref{tab:senTab} gives an overview of the distribution of the sentiment variables across the entire data set.

\begin{table}[ht]
\centering
\caption{Distribution parameters of message sentiments.} 
\label{tab:senTab}
\begin{tabular}{lrrrr}
  \toprule
Variable & Min & Mean & Max & SD \\ 
  \midrule
Polarity & 0.04 & 16.13 & 111.39 & 22.04 \\ 
  Anger & 1.47 & 1.53 & 7.34 & 0.59 \\ 
  Disgust & 3.09 & 3.44 & 7.34 & 1.17 \\ 
  Fear & 2.07 & 2.09 & 7.34 & 0.37 \\ 
  Joy & 1.03 & 2.89 & 19.97 & 3.45 \\ 
  Sadness & 1.73 & 2.12 & 7.34 & 1.43 \\ 
  Surprise & 2.79 & 3.17 & 11.89 & 1.31 \\ 
   \bottomrule
\end{tabular}
\end{table}

\begin{figure}
{\centering \includegraphics{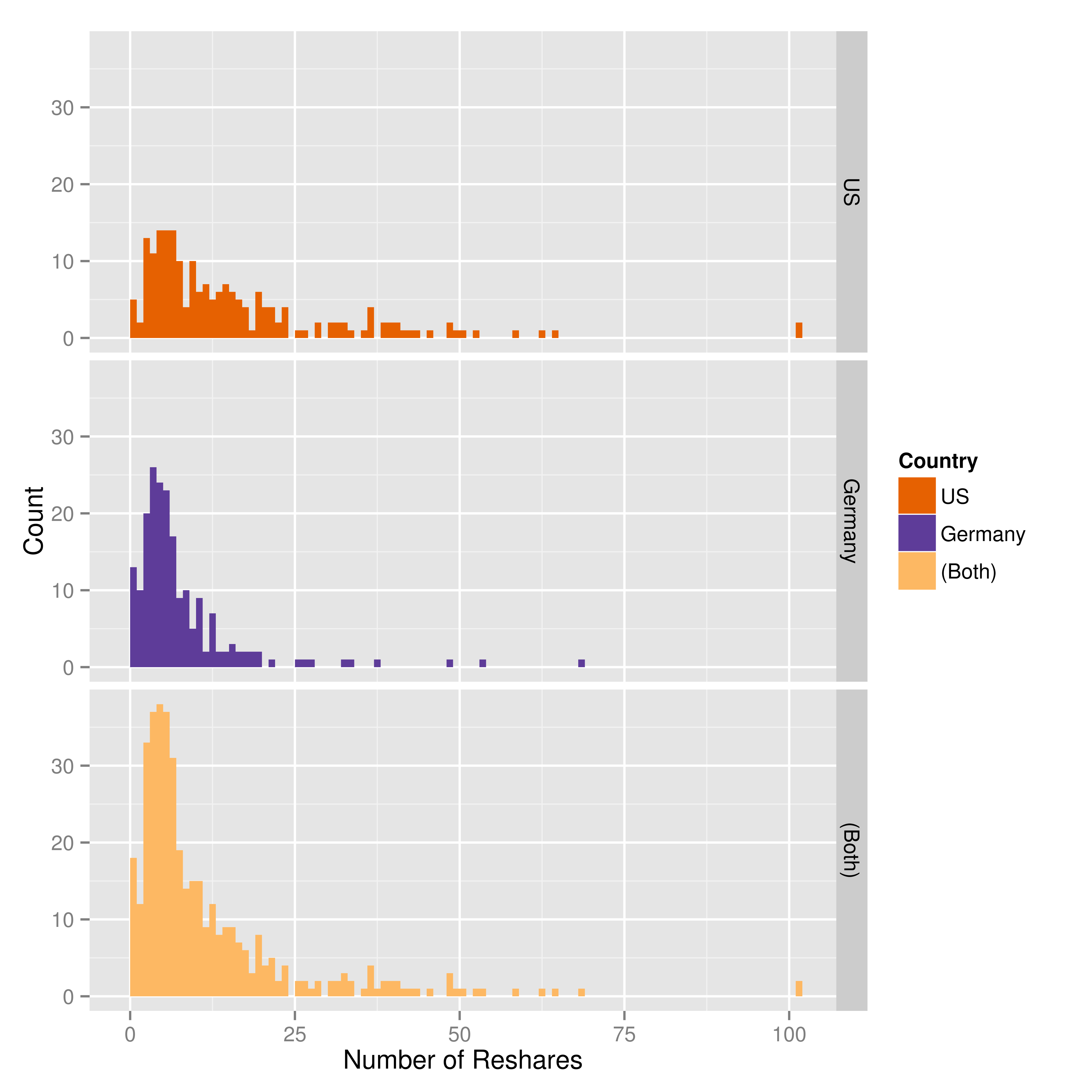} 

}

\caption[Comparing the absolute number of reshares accross countries]{Comparing the absolute number of reshares accross countries.\label{fig:nR-diffs-abs}}
\end{figure}

There is a considerable difference in the frequency and volume of messages being reshared between both countries, as is demonstrated by Fig.~\ref{fig:nR-diffs-abs}. In order to explain -- at least partially -- these differences, we use negative binomial models with a logistic link function.

The negative binomial model is an extension to the more familiar Poisson regression model. The latter frequently suffers from overdispersion, that is its sample variance exceeds its sample mean. It is possible to justify the appropriateness of a negative binomial model over a Poisson regression model, using a $\chi^2$-test \cite{mass}.

In order to model the reshare rate of a message (as opposed to raw counts of reshares), one has to take into account the potential exposure of a message. This is directly affected by the number of followers a G+ page has and the age of the message: the more followers to a page and the longer the message has been online, the more people are likely to have come across it. These measurement windows or base references are included as log offset terms in the models \cite{mass}. 

For developing the model, we combine model enhancement with backward selection based on AIC. We start out with a simple model using the classical covariates as described above and optimizing it using backward selection (M1). In a next step, we introduce \emph{Country} (US being the reference category) as an interaction variable and employ backward selection again (M2). Finally, we include the variables from our sentiment analysis (M3). All models were tested for the appropriateness of a negative binomial specification using aforementioned $\chi^2$-test.

In all models, the logarithm of message age is used as an offset to model exposure. For models that do not contain the variable \emph{country}, the logarithm of the number of followers of a G+ page is included as an additional offset.

\begin{table}
\caption{\label{tab:modSum}Summary of model components.}
\centering
\begin{tabular}{ll}
\toprule
\textbf{Model} & \textbf{Components} \\
\midrule
M1 & Classical covariates\\
M2 & M1 \& interacting with country\\
M3 & M2 \& sentiment variables (interacting with country)\\
\bottomrule
\end{tabular}
\end{table}

\section{Results}

In this section we present the results of our analysis. Tables \ref{tab:m1} through \ref{tab:m3} list the estimated parameters, their standard errors, and p-values for all three models. Note however, that the p-values are given for reference purposes only. We are aware that we are not working with a random sample here and p-values are therefore ultimately meaningless. This is not problematic, as we are following a data mining approach (as opposed to a classical inferential statistics approach) and are interested in real observations and real differences and not estimates that generalize towards a larger population.

\begin{table}[ht]
\centering
\caption{Estimated parameters for model M1.} 
\label{tab:m1}
\begin{tabular}{lrrr}
  \toprule
Parameter & Estimate & Std.Error & p.value \\ 
  \midrule
Intercept & -13.42 & 0.08 & 0.00 \\ 
  Night & 0.52 & 0.23 & 0.02 \\ 
  Tuesday/Saturday & 0.50 & 0.15 & 0.00 \\ 
   \bottomrule
\end{tabular}
\end{table}

In a first step we sought to assess the classical covariates established in the literature. Stepwise selection found that only time- and date-related covariates are important parameters at this step. When allowing the stepwise selection algorithm to choose from interaction terms with the country of the G+ page, the daytime related term in the model becomes dependent on the country. The change in sign is a clear indication for the plausibility of the additional interaction. Model fit increased as well with $AIC_{M2} = 3004$ compared to $AIC_{M1} = 3113$. 

\begin{table}[ht]
\centering
\caption{Estimated parameters for model M2.} 
\label{tab:m2}
\begin{tabular}{lrrr}
  \toprule
Parameter & Estimate & Std.Error & p.value \\ 
  \midrule
Intercept & -1.30 & 0.11 & 0.00 \\ 
  Evening/Night & 1.02 & 0.25 & 0.00 \\ 
  Country & -1.26 & 0.15 & 0.00 \\ 
  Tuesday/Saturday & 0.71 & 0.14 & 0.00 \\ 
  Evening/Night x Country & -0.55 & 0.31 & 0.08 \\ 
   \bottomrule
\end{tabular}
\end{table}

The addition of the message sentiment variables improves the fit only marginally: $AIC_{M3} = 2994$. However, two sentiment variables' effects on the reshare rate differ across countries: Anger and Surprise.

\begin{table}[ht]
\centering
\caption{Estimated parameters for model M3.} 
\label{tab:m3}
\begin{tabular}{lrrr}
  \toprule
Parameter & Estimate & Std.Error & p.value \\ 
  \midrule
Intercept & -2.37 & 0.40 & 0.00 \\ 
  Evening/Night & 1.07 & 0.24 & 0.00 \\ 
  Country & -0.01 & 0.46 & 0.98 \\ 
  Tuesday/Saturday & 0.72 & 0.14 & 0.00 \\ 
  Disgust & 0.11 & 0.06 & 0.06 \\ 
  Anger & 0.44 & 0.21 & 0.04 \\ 
  Surprise & 0.01 & 0.08 & 0.92 \\ 
  Evening/Night x Country & -0.70 & 0.31 & 0.02 \\ 
  Anger x Country & -0.44 & 0.24 & 0.07 \\ 
  Surprise x Country & -0.19 & 0.10 & 0.06 \\ 
   \bottomrule
\end{tabular}
\end{table}

\section{Discussion}
These models allow for a number of insights. Foremost, we find in line with most literature on message propagation in social networks, that time of day has a highly significant effect on the number of rebroadcasts a message will receive. It is noteworthy, however, that the size of the effect changes with the country of the observation. Apparently, German G+ pages don't benefit from nighttime posts as much as Americans do. Table \ref{tab:eff.tod} presents the marginal effects of posting time across country.

\begin{table}[ht]
\centering
\caption{Marginal effect of a daytime post on expected reshare count across countries.} 
\label{tab:eff.tod}
\begin{tabular}{lrr}
  \hline
Country & Daytime & Nighttime \\ 
  \hline
Germany & 7.42 & 10.73 \\ 
  US & 27.15 & 79.24 \\ 
   \hline
\end{tabular}
\end{table}

Day of week also has a very significant influence here. Messages sent out on Tuesdays or Saturdays will receive increased attention. This effect is consistent across all models and does not depend on country.

\begin{figure}[]

{\centering \includegraphics{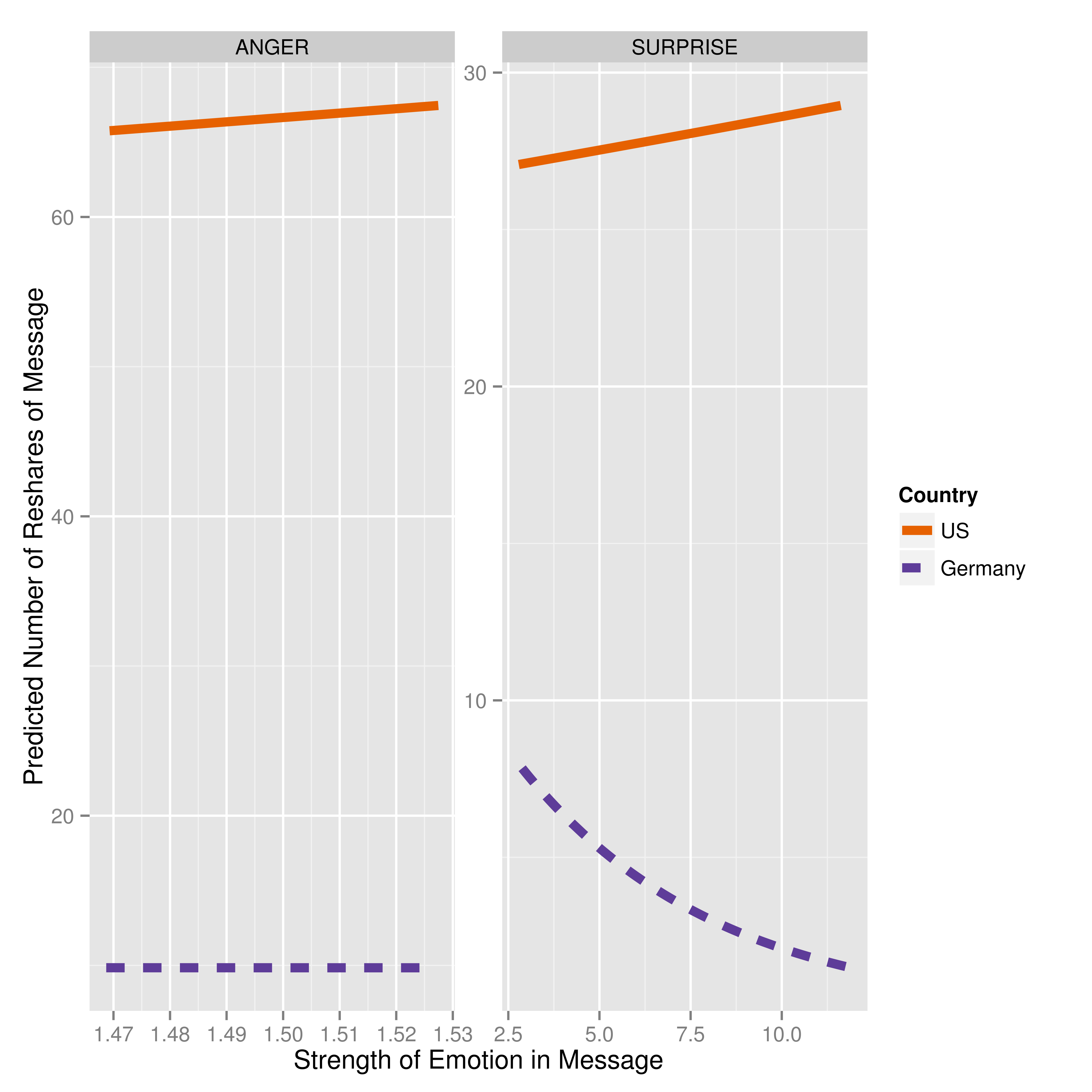} 

}

\caption[Marginal effects of Anger and Surprise on expected reshare counts for US and German G+ pages]{Marginal effects of Anger and Surprise on expected reshare counts for US and German G+ pages.\label{fig:eff.plot.both}}
\end{figure}

Finally, there is another effect here that is dependent on culture: the role of emotion, more precisely of Anger and Surprise. There is a rather strong effect testifying to the German distaste of surprises. The expected count of reshares for a message increases with its notion of surprise in the US, while in Germany surprises slightly decrease these chances. A similar observation holds for Anger. Fig.~\ref{fig:eff.plot.both} exhibits the sizes of these effects.

In light of the differences in culture observed by Hofstede, we found evidence for message traits that are consistent across diverse cultures differing in the individuality and uncertainty domains. This is chiefly the preference of certain weekdays and the perhaps obvious rejection of disgusting emotions embedded in advertising messages. 

Other message and communication properties, time of day when the message was sent and angry or surprising sentiments, differ between the two cultures. Consumers in the US seem to prefer messages that are being sent out at nighttime, while their German counterparts are rather indifferent in that regard. This might be a sign for America's highly valued individualism and work ethics that would postpone past-time activities like checking Google+ pages to the nighttime.

German distaste of surprises and angry messages can as well be explained in terms of Hofstede's cultural dimensions: Uncertainty is something that is immanent to surprises and also anger poses certain risks. Germany's high level of uncertainty avoidance might explain the increased resharing of messages exhibiting neither anger nor surprise. 

\section{Conclusion}

In light of these findings, it becomes clear that marketers must consider the culture of their audience even in an online arena. Alas, as Kitchen puts it:

\begin{quote}
``Social media can be a proxy for consumer ethnography, the
anthropological approach of understanding a culture by
becoming part of it, because they provide virtual access to an
often unguarded engagement with quite intimate aspects of
consumer experience.'' \cite[p.~36]{kitchen2013dominant}
\end{quote}

And who would not be cautious when treading with marketing intentions in intimate places? Cross-cultural communication might be hindered by a number of factors: ethnocentrism (i.e. ignoring other cultures), parochialism (focusing too much on local peculiarities) and stereotyping (because reality can be to complex to model) \cite{hurn2013cross}.

While it has become clear in the last few decades that ethnocentrism is a deadly sin for marketers and parochialism equally hinders effective campaigns, stereotyping is still all too familiar in marketing endeavors. After all, segmented target \emph{groups} are defined by broad \emph{averages} that marketers seek to please. And indeed, to avoid stereotyping, one needs not only to know the culture of a market, but to respect and adapt to individual consumers, wherever possible \cite{torelli2013globalization}. 

We believe that this adaption is easier to be had than previously thought: with the massive amounts of data available from social media platforms, marketers can seek to understand consumers more directly, away from cultural considerations, and just watch them forwarding and endorsing messages. It is this observation that empowers us to learn from consumers what they want. 

\bibliographystyle{plainnat}
\bibliography{hw-cult.bib}

\end{document}